%% file: main.tex
\documentclass[10pt,journal,compsoc]{IEEEtran}
\usepackage{tikz}
\usepackage{amsmath}

\usepackage{stfloats}
\usepackage{filecontents}
\usepackage{verbatim}
\usepackage{enumerate}

\usepackage{epsfig} 
\usepackage{setspace}
\usepackage{balance}
\usepackage{algorithm}
\usepackage{algorithmic}
\usepackage{amsfonts}
\usepackage{authblk}
\usepackage{epsfig}  
\usepackage{amssymb}
\usepackage[normalem]{ulem}
\usepackage{amscd}
\usepackage{amsmath} 
\usepackage{microtype}
\usepackage{array}
\usepackage{chngpage}
\usepackage{graphicx}
\usepackage{epsfig}
\usepackage{multirow}
\usepackage{caption}
\usepackage{listings}
\usepackage{subfigure}
\usepackage{adjustbox}
\usepackage{ulem}
\usepackage{lipsum}
\usepackage{float}
\usepackage{pifont}
\usepackage{bm}
\usepackage{color,xcolor}
\usepackage{booktabs} 
\usepackage{verbatim}
\usepackage{amssymb}
\usepackage{pifont}
\usepackage{array,multirow}
\usepackage{makecell}
\usepackage{tikz}
\usepackage{CJK}
\usepackage{float}
\usepackage{threeparttable}
\usepackage{colortbl}
\usepackage{titlesec}
\usepackage{cleveref}
\usepackage[utf8]{inputenc}
\usepackage{algorithmic}
\crefname{section}{§}{§§}
\Crefname{section}{§}{§§}
\usepackage{fancyhdr}

\newcommand*\circled[1]{\tikz[baseline=(char.base)]{
            \node[shape=circle,fill,inner sep=0.6pt] (char) {\textcolor{white}{#1}};}}

\newcommand{\fnam}{\text{\textit{SEAM}}}
\newcommand{\fname}{\text{\textit{SEAM} }}
\newcommand{\mnam}{\text{\textit{BinTran}}}
\newcommand{\mname}{\text{\textit{BinTran} }}
\newcommand{\inam}{\text{\textit{SeamCode}}}
\newcommand{\iname}{\text{\textit{SeamCode} }}

\newcommand{\ignore}[1]{}
\newcommand{\todo}[1]{\textcolor{red}{(#1)}}
\newcommand\cnum[1]{{\textcolor{black}{#1}}}

\newcommand\kai[1]{{\textcolor{blue}{Kai: #1}}}

\ifCLASSOPTIONcompsoc
  \usepackage[nocompress]{cite}
\else
  \usepackage{cite}
\fi
\ifCLASSINFOpdf
\else
\fi
\begin{document}
%
\title{Semantics-Recovering Decompilation through Neural Machine Translation}
%
%
%
%

\author{Ruigang Liang,
        Ying Cao,
        Peiwei Hu,
        Jinwen He,
        Yingjun Zhang,
        and~Kai Chen
\thanks{Ruigang Liang, Ying Cao, Peiwei Hu, Jinwen He, and Kai Chen are with the State Key Laboratory of Information Security, Institute of Information Engineering, Chinese Academy of Sciences, Beijing 100093, China, and also with the School of CyberSecurity, University of Chinese Academy of Sciences, Beijing 100093, China, e-mail: (liangruigang, caoying, hupeiwei, hejinwen, chenkai)@iie.ac.cn.}
\thanks{Yingjun Zhang is with the Institute of Software, Chinese Academy of Sciences, Beijing 100190, China (e-mail: yjzhang@tca.iscas.ac.cn).}
\thanks{Manuscript received June 28, 2021.}}

%
%

\markboth{}%
{Shell \MakeLowercase{\textit{et al.}}: Bare Demo of IEEEtran.cls for Computer Society Journals}
%



\IEEEtitleabstractindextext{%
\begin{abstract}
Decompilation transforms low-level program languages (PL) (e.g., binary code) into high-level PLs (e.g., \textit{C/C++}). It has been widely used when analysts perform security analysis on software (systems) whose source code is unavailable, such as vulnerability search and malware analysis. However, current decompilation tools usually need lots of experts' efforts, even for years, to generate the rules for decompilation, which also requires long-term maintenance as the syntax of high-level PL or low-level PL changes. Also, an ideal decompiler should concisely generate high-level PL with similar functionality to the source low-level PL and semantic information (e.g., meaningful variable names), just like human-written code. Unfortunately, existing manually-defined rule-based decompilation techniques only functionally restore the low-level PL to a similar high-level PL and are still powerless to recover semantic information. 
In this paper, we propose a novel neural decompilation approach to translate low-level PL into accurate and user-friendly high-level PL, effectively improving its readability and understandability. Furthermore, we implement the proposed approaches called {\it SEAM}. Evaluations on four real-world applications show that {\it SEAM} has an average accuracy of 94.41\%, which is much better than prior neural machine translation (NMT) models. Finally, we evaluate the effectiveness of semantic information recovery through a questionnaire survey, and the average accuracy is 92.64\%, which is comparable or superior to the state-of-the-art compilers.
\end{abstract}

\begin{IEEEkeywords}
Neural decompilation, neural machine translation, semantics information recovery.
\end{IEEEkeywords}}

\maketitle

\IEEEdisplaynontitleabstractindextext

\IEEEpeerreviewmaketitle

\input{Introduction}

\input{Background}

\input{Motivation}

\input{Approach}

\input{Implementation}

\input{Evaluation}

\input{Discussion}

\input{Related_work}

\input{Conclusion}

\bibliographystyle{IEEEtran}
\bibliography{reference.bib}

\end{document}

%% file: Introduction.tex
\IEEEraisesectionheading{\section{Introduction}\label{sec:introduction}}

\IEEEPARstart{D}{ecompilation} comes into being along with the emergence of compilation and evolves with compilation technology. The goal of decompilation is to transform a low-level program language (PL) generated by a compiler (e.g., assembly language) into a high-level PL (e.g., C/C++) that is functionally equivalent and friendly to read. Decompilation serves as the core technology in software reverse engineering, which has been widely used in software security analysis, such as vulnerability discovery and malicious code detection. However, it is non-trivial to build an effective decompiler that generates human-friendly code for program analysis.



State of the art in recent enormously depends on manually-defined rules, such as Hex-Rays~\cite{Hex-Rays}, Retdec~\cite{kvroustek2017retdec} and Ghidra~\cite{Ghidra}. It makes them suffer from the following shortages: (1) Long development cycle, mainly relying on expert experience to design a specific PL. For example, as we know, developing the RetDec~\cite{kvroustek2017retdec} requires about \cnum{7} years of experts' hard work~\cite{avast-retargetable, katz2019towards}. (2) Lack of scalability, requiring significant engineering overhead when inserting new rules or decompiling a new high-level PL. (3) Poor readability of the decompiled high-level PL. Merely functionally restoring the low-level target PL to a similar high-level PL, without effective semantic information recovery. An example of Hex-rays (the most famous commercial decompilation tool) decompilation is shown in List \ref{lst:example-HexRays}, specific analysis is in Section~\ref{sec:Motivation}. It recovers variables and function names as number strings without semantics (e.g., variable \textit{v5} in line \cnum{4}), which is even hard to understand by an expert.

Inspired by the success of natural language processing (NLP)~\cite{DBLP:journals/corr/abs-2003-11080} and NMT~\cite{vaswani2017attention}, our idea is to translate assembly code to source code by NMT, which performs excellent in natural language translation (e.g., English to Chinese). Fu et al.~\cite{fu2019coda} presented a neural decompilation framework that can recover semantic information for some simple operations (e.g., arithmetic), but it remains unsolved for recovering complex operations and hardware instruction set architecture (ISA), such as x86-64. An ideal decompiler should correctly recover low-level PL code to high-level PL code with human-friendly semantic information. Unfortunately, the absence of semantic information in low-level PL causes the following three main challenges.




\vspace {3pt}\noindent\textbf{Challenges}.
\textbf{C1:} The mapping rule is unclear. Unlike natural languages, there are asymmetries between low-level PL and high-level PL, even if they have the same functionalities. Considering that an NMT model is mainly used to translate two languages with few information asymmetries, it is not feasible to directly use the model to translate low-level PL code to high-level PL code. Such information asymmetry is fundamentally caused by the difference between the high-level PL and the instruction sets of the CPU (expressed by low-level PL). In the compilation process, the semantic information unrelated to the instruction sets is removed by the compiler. For example, a symbol table is a data structure generated by the compiler during compilation, mainly used to store identifiers and their declaration or usage information in high-level PL. Symbol tables are not required for binary executables and are often removed by tools such as ``strip''. Therefore, assembly code does not contain such symbol information, such as variable types and variable names. However, the missed information for helping developers understand the code cannot be recovered solely from the low-level PL code, making NMT unable to fully recover the original high-level PL code. For example, a recent study~\cite{katz2018using} directly uses the NMT model recurrent neural networks (RNN) for decompilation, which fails to recover semantic information such as variable name and function name. Also, to make the low-level PL code run on the CPU, it contains more redundant information (such as jump address, function call address), which should be removed from the translation. Moreover, to improve code execution efficiency, instructions in the low-level PL code may also be optimized. Some relatively simple operations for the high-level PL code are converted into compound operations suitable for machine execution. For example, for the division operation $var2=var1/3$ of the high-level PL, the low-level PL converts it to a low-level PL in Listing~\ref{lst:An example of division operation of low-level PL} (note that there is no \textit{div} instruction here), and the method of conversion changes as the divisor changes. Making the neural model learn the mapping rules between high-level PL and low-level PL turns out to be nontrivial.

\begin{lstlisting}[language=c,frame=trBL,caption=An example of division operation of low-level PL,abovecaptionskip=10pt,belowcaptionskip=0pt,captionpos=b, label={lst:An example of division operation of low-level PL},
  keywordstyle=\bfseries\color{green!40!black},
  commentstyle=\itshape\color{purple!40!black},
  identifierstyle=\color{blue},
  stringstyle=\color{orange},
  basicstyle=\footnotesize\ttfamily,
  breaklines=true,
  numbers=left,
  numbersep=-5pt,
  stepnumber=1,]
  mov     eax, DWORD PTR [rbp-4]
  movsx   rdx, eax
  imul    rdx, rdx, 1431655766
  shr     rdx, 32
  sar     eax, 31
  mov     ecx, edx
  sub     ecx, eax
  mov     eax, ecx
  mov     DWORD PTR [rbp-8], eax
\end{lstlisting}


\vspace {3pt}\noindent\textbf{C2:} 
Compared to natural language, the syntax requirements of PL are more stringent. For example, when translating the expression \texttt{(a + b) * c}, the position information of characters is required to be strictly accurate, such as \textit{+} must be between \textit{a} and \textit{b}, \textit{*} must be between \textit{)} and \textit{c}. By contrast, synonyms (e.g. \textit{dog} and \textit{puppy}) and pronoun (e.g. \textit{he}, \textit{she}) exist in natural language. A word can be replaced with its synonym in the same context, and pronouns can be used to refer to people or things that occur in the context above. However, PL requires that the name of the same entity must be strictly identical. Otherwise, it will lead to syntax or function errors.

\vspace {3pt}\noindent\textbf{C3:} The identifier's semantic information cannot get from the low-level PL, which requires additional information. The semantic information associated with the PL code is stored separately in the symbol table during the compilation process, and the resulting low-level PL does not have semantic information itself. Before releasing the software, developers often remove debugging information to reduce the file size. So the recovery of semantic names for identifiers has always been a problem in decompilation research. In the real-world programming process, technicians tend to follow specific programming specifications, such as Google C++ Style Guide\footnote{https://google.github.io/styleguide/cppguide.html}, or to name functions and variables according to the program's functionality or some conventions. For example, functions with dichotomous lookup are usually named with binary-related words, and counters in for loops are usually named \texttt{i}, \texttt{j}, \texttt{k}, and so on. The names of variables with semantic information are usually beneficial for the understanding of PL. How to recover them is still a challenge for NMT.

\vspace {3pt}\noindent\textbf{Our approach}.
In this paper, we overcome the challenges mentioned above and design a neural decompilation framework, called \fnam\footnote{{\it SEAM} (Semantics-Recovering Decompilation)}, capable of translating low-level PL code to high-level PL code and recovering semantic information about identifiers. The core idea of \fname has two main points. The first is to compensate for the information asymmetry that exists between the low-level PL and high-level PL by serializing the representation of the high-level PL (similar to the representation of the low-level PL) and the statement normalization operation so that the NMT model can accurately learn the transformation rules between the high-level PL and the low-level PL. The second point is to compensate for the lack of semantic information in the decompiled high-level PL by introducing additional information and to recover its semantic information by renaming identifiers such as function and variable names. \fname improves the accuracy of the semantic and functional structure of the decompiled high-level PL and significantly improves its readability and comprehensibility.


\fname works in three phases: code regularization, translation through neural translation model, and semantic information recovery. 
In the first phase, \fname is committed to the regularization of PL, aiming to make the model learn the decompilation rules between the high-level PL and the low-level PL well. 
To compensate for the information asymmetry between the low-level PL and the high-level PL as mentioned in C1, we propose a \iname generation approach. We design a new intermediate language \iname for high-level PL, transforming the hierarchical high-level PL code into a serialized representation similar to the low-level PL code. On the one hand, \iname removes the syntax structure of high-level PL code (addressing C2), effectively retains the semantic information, and can be directly converted into high-level PL code. On the other hand, it has a similar expression form as the low-level PL. Further, to make up for the asymmetry in identifier semantics between the low-level PL and high-level PL (addressing C1), we propose a statement canonicalization approach where identifiers of the PL are uniformly substituted in units of type to reduce their adverse effects on model training. Using the regularized processed high-level PL (\inam) and the low-level PL pairs as inputs effectively reduce the difficulty of modeling learning transformation rules.

In the second phase, after preprocessing the PL dataset with the code regularization, \fname utilizes the dataset as the training set for the neural translation model that we design based on self-attention mechanism, called \mnam, accurately learn the conversion rules between low-level PL and high-level PL.
In order to avoid the function (code) in the real-world low-level PL is too long (exceeding the maximum length limit of the model input) to affects the accuracy of the model, we propose a low-level PL segmentation approach that constructs PL code in units of high-level PL lines, which effectively addresses the effect of the model on input length constraints. 


In the third phase, \fname builds a semantic information recovery model based on the idea of image caption generation~\cite{vinyals2015show} in NLP, aiming to recover the semantics of decompiled high-level PL (addressing C3) through the provision of additional information. The model generates identifier names with semantic information based on the functionality of the PL and then embeds them into the generated high-level PL sketch generated in the second phrase, achieving accurate recovery of decompiled PL semantic information and improving its readability.

We implement \fname on the base of the self-attention mechanism and further evaluate the performance using real-world projects. The results show that \fname achieves an average accuracy of \cnum{94.41}\% on \cnum{4} real-word projects. Compared to Hex-Rays~\cite{Hex-Rays}, the most famous commercial rule-based decompiler, \fname can accurately recover identifier semantics. Based on our human subject study of \cnum{42} developers from big companies such as Google and Tencent, 97.96\% of the identifiers are meaningful. The results demonstrate that the output of \fname recovers both functionality and semantics, significantly improving the readability and comprehensibility.


\vspace{5pt}\noindent\textbf{Contributions}. Our contributions are outlined as follows:

\vspace {5pt}\noindent$\bullet$\space\textit{New technique.} We design a novel neural decompilation technique that translates low-level PL into accurate and user-friendly high-level PL. To the best of our knowledge, \fname is the first neural decompilation framework that can accurately recover semantic information about identifiers. \fname addresses several key challenges that prior research has not effectively overcome, including building an intermediate language that can serialize the representation of high-level PL, accurately recovering abstract syntax tree (AST) sketch for high-level PL using the self-attention mechanism, recovering PL identifier semantics through the functions of code. Evaluation results show that \fname is comparable or superior to the state-of-the-art compilation tools.

\vspace {5pt}\noindent$\bullet$\space\textit{New Understanding}. Our study demonstrates that neural translation techniques for natural languages are also valuable for PL's decompilation task, as long as the information asymmetry between low-level PL and high-level PL is appropriately handled. We design an intermediate representation \iname to address such asymmetry and introduce attention-based techniques to implement \fnam, which performs well in translation and helps to provide an interpretable understanding of the translation.

%% file: Background.tex
\section{Background}
\label{sec:Background}

This section gives a brief introduction to decompilation technology and recent studies utilizing deep learning to improve the semantic and syntactic correctness of decompiled code.

\subsection{Decompilation}

Decompilation is a technique that transforms a compiled executable program or low-level PL with intermediate representation into a functionally equivalent and easy-to-read high-level PL~\cite{van2007static}. \ignore{It is the inverse process of compilation.} The first mature decompiler appeared about a decade later than compilers~\cite{10.5555/1177220}, which implied its difficulty. One of the main difficulties lies in the missing information related to the original program during compilation. The binary code in a highly abstract machine language generated by compilers lacks much semantic information. There are many practical applications for decompilation. On the one hand, decompilation can be used to understand the inherent mechanisms of low-level PL (like binary code, assembly, and intermediate language), discover software vulnerabilities, analyze malicious code, or perform verification and program comparison~\cite{van2007static}. On the other hand, decompilation provides technical support for source-level analysis and optimization tools, making it relatively easy to port programs to new hardware architectures or operating systems when the source code is available and can be compiled to a new environment.

Decompilation research can be broadly categorized as either (1) rule-based approaches~\cite{Hex-Rays, brumley2013native, kvroustek2017retdec, Ghidra} or (2) more recent NMT-based approaches~\cite{katz2018using, katz2019towards, fu2019coda,liang2021neutron}. Rule-based approaches for decompilation started to gain popularity with the advent of Hex-Rays~\cite{Hex-Rays}, a commercial native processor code to C-like code decompiler. The Hex-Rays becomes an essential tool for reverse engineers. A brief history of the related NMT-based approach starts with Katz et al.~\cite{katz2018using}. In their work, they draw on insights and methods from machine learning and NLP and categorize decompilation as a language translation problem, which also has brought a new idea to the research of decompilation. However, even the state-of-the-art decompilers (either rule-based or NMT-based approaches) share similar bottlenecks. Most of the difficulties stem from the need to recover some debug information that is not explicitly provided in low-level PL, such as variable and function name semantics, data types, and the correct code optimized by compilers. In this paper, we design a framework named \fnam, together with a Neural Translation Model \mname to achieve this goal without artificially specified rules. To the best of our knowledge, this is the first neural decompilation framework that can accurately recover semantic information.

\subsection{Deep Learning on Decompilation}

Deep learning applied to decompilation in reverse engineering and security research offers new insights into solving challenges in traditional rule-based research. The core research direction is to convert the decompilation of code into translation problems between different PL, drawing on the current increasingly mature NMT approach~\cite{bahdanau2014neural, sutskever2014sequence, wu2016google, vaswani2017attention}. Bahdanau et al.~\cite{bahdanau2014neural} proposed an encoder-decoder NMT model architecture, which then becomes the mainstream architecture to model translation. In this architecture, the encoder and decoder can be considered as two separate models, with the encoder (model) taking care of encoding the neutral input language into a vector representation and the decoder (model) responsible for decoding it into the target language~\cite{lacomis2019dire}. Encoders and decoders typically use RNN due to the high-dimensional hidden state and nonlinear dynamics of recurrent neural networks, remembering and processing previous data, and are well suited for serialized information such as natural language. The decompilation architecture of TraFix~\cite{katz2019towards} and Coda~\cite{fu2019coda} are based on the encoder-decoder NMT model.

However, RNN has some drawbacks. First, due to its serialized network architecture design, it is impossible to parallelize, resulting in slower and more costly training. Second, for long-distance or hierarchical data, RNN has difficulty establishing its dependencies. The attention mechanism~\cite{bahdanau2014neural} is added to encoder-decoder NMT architecture for overcoming these problems, whose core role is to learn and mark the importance of each word from the sequence and map it to the output token. Vaswani et al.~\cite{vaswani2017attention} proposed a network architecture Transformer that based solely on attention mechanisms, which considered not only the dependency relationship between the input and output sequence words but also the correlation relationship between words inside the input or output sequence itself. We apply the self-attention mechanism to decompilation and design \mname model based on Transformer to make it suitable for the PL translation.

%% file: Motivation.tex
\section{Motivation}
\label{sec:Motivation}

As mentioned above, the goal of the decompiler is to perform a series of analyses and recover from binary or intermediate low-level PL code the functionally equivalent high-level PL code when the high-level PL is not available, and to provide technical support for vulnerability, malware discovery, verification, and other program security analysis. Existing mainstream decompilers are mainly rule-based implementations with long development cycles (according to data published by Avast, development of RetDec~\cite{kvroustek2017retdec} require 24 developers to take more than 7 years~\cite{katz2019towards}). The bottleneck is poor scalability (when new rules need to be inserted or new languages need to be decompiled at a high cost) and over-reliance on expert experience. Further, the translation results of these decompilers are unsatisfactory. They only functionally restore the low-level target PL to a similar high-level PL without effective recovery of semantic information or user-defined datatypes. The readability of high-level PL is still low. Only experts can accurately understand the underlying logic. Since the semantic information related to the low-level PL is stored separately in the symbol table during the compilation process, the generated low-level PL does not have semantic information. The recovery of semantic information such as names of identifiers has been challenging in decompilation research.

If there is an approach that can complete the learning and extraction of PL rules through intelligent learning mechanism, and achieve accurate recovery of PL functions and semantic information, assisting analysts to complete accurate understanding and analysis of the target low-level PL intrinsic mechanism, further breaking the over-reliance on decompilation techniques for rule definition. In this way, the scalability of the decompiler can be improved, the development cycle can be shortened, and the readability of decompiled programs can be guaranteed. Inspired by the successful application of deep learning in the field of machine translation~\cite{devlin-etal-2019-bert}, we are wondering whether deep learning could be applied to the decompilation of PL. Through the process of decompilation and machine translation, we find similarities between them: they both translate (decompile) one language into another through specific rules, and there are functional similarities between the input (source language) and the output (target language). However, there are also essential differences between them, such as information asymmetry, grammar restrictions, language structure, translation rules, length of corresponding statements, etc. As mentioned above, all of these make the decompilation process very challenging (see Section~\ref{sec:introduction}).

\begin{lstlisting}[language=c,frame=trBL,caption=An example of decompilation utilizing Hex-Rays,abovecaptionskip=10pt,belowcaptionskip=0pt,captionpos=b, label={lst:example-HexRays},
  keywordstyle=\bfseries\color{green!40!black},
  commentstyle=\itshape\color{purple!40!black},
  identifierstyle=\color{blue},
  stringstyle=\color{orange},
  basicstyle=\footnotesize\ttfamily,
  breaklines=true,
  numbers=left,
  numbersep=-5pt,
  stepnumber=1,]
  int__cdecl main(int argc, const char **argv,const char **envp)
  {
    __int64 v3; // rdx
    int v5; // [rsp+Ch] [rbp-14h]
    int v6; // [rsp+10h] [rbp-10h]
    int v7; // [rsp+14h] [rbp-Ch]
    unsigned __int64 v8; // [rsp+18h] [rbp-8h]
    
    v8 = __readfsqword(0x28u);
    __isoc99_scanf(&unk_4006D4, &v5, envp);
    __isoc99_scanf(&unk_4006D4, &v6, v3);
    v7 = v5 + v6;
    if ( v5 + v6 == (_DWORD)&qword_400400 ) 
        printf("You Win!");
    return 0;
  }

\end{lstlisting}

\begin{lstlisting}[language=c,frame=trBL,caption=Source code for the Hex-Rays decompilation example,abovecaptionskip=10pt,belowcaptionskip=0pt,captionpos=b, label={lst:Source code for the Hex-Rays decompilation example},
  keywordstyle=\bfseries\color{green!40!black},
  commentstyle=\itshape\color{purple!40!black},
  identifierstyle=\color{blue},
  stringstyle=\color{orange},
  basicstyle=\footnotesize\ttfamily,
  breaklines=true,
  numbers=left,
  numbersep=-5pt,
  stepnumber=1,]
  #include <studio.h>
  #include<stdlib.h>

  int main() {
      int num1, num2, sum;
      scanf("%d", &num1);
      scanf("%d", &num2);
      sum = num1+num2;
      if (sum == 0x400400)
          printf("You Win!");
  }
\end{lstlisting}

\vspace {3pt}\noindent\textbf{Example}. List \ref{lst:example-HexRays} gives an example of decompilation utilizing Hex-Rays. The corresponding source code can be found in  List \ref{lst:Source code for the Hex-Rays decompilation example}. None of the identifiers in the code segments for high-level PL has semantic information. Even worse, the decompiled result is incorrect. It can be seen that line 9 of the source code in list \ref{lst:Source code for the Hex-Rays decompilation example} is a conditional judgment statement that tests whether the value of the variable \texttt{sum} is equal to \texttt{0x400400}. If it is equal to this value, the \texttt{printf} statement in the \texttt{if} condition is executed. The variable \texttt{sum} is just a normal int-type variable whose value also depends on the value read by \texttt{scanf} on line 6 and line 7. However, from line 9 of list \ref{lst:Source code for the Hex-Rays decompilation example}, it can be seen that Hex-Rays identified the value \texttt{0x400400} as an address and underwent a series of processes so that the decompiled code works incorrectly. We found that Hex-Rays attempt to utilize optimization to translate simpler code statements during low-level PL decompilation, but because of its inaccurate identification of addresses, this optimization instead leads to many errors, which interferes with the accurate analysis of the target program. Our idea is to treat low-level PL and high-level PL as two different language forms and convert the low-level PL's decompilation into translation problems between two different PLs.

Different from previous researches on decompilation using deep learning~\cite{katz2018using}, we do not directly apply the NMT model to PL decompilation~\cite{katz2019towards}, nor do we construct PL decompilation methods through the nesting of multiple models~\cite{fu2019coda}. Our approach focuses on the regularization of PL data to compensate for the semantic asymmetry between low-level PL and high-level PL so that the NMT model can accurately learn the decompilation rules. By providing additional information to help recover semantic information of the decompiled high-level PL, we improve the readability and comprehensibility of the decompiled high-level PL.


%% file: Approach.tex
\section{Approach}
\label{sec:Approach}

\begin{figure*}
\centering
\epsfig{figure=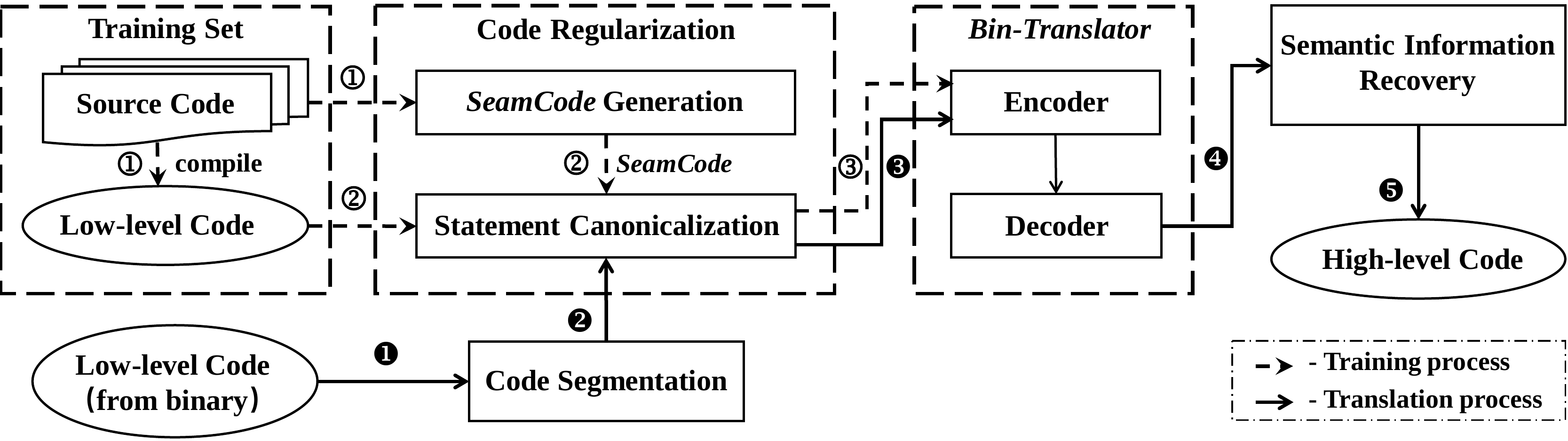, width=0.85\textwidth} 
\caption{\textbf{Overview of \fnam}}
\vspace{-10pt}
\label{pic:framework}
\end{figure*}

We propose the design of \fnam, a neural approach to decompile low-level PL to C-like high-level PL automatically. Such a framework can improve the semantic and syntactic accuracy of decompiled high-level PL and significantly enhance its readability. Below we elaborate on the details of \fnam.

\subsection{Overview}
\label{subsec:overview}

Figure~\ref{pic:framework} shows the overview of \fnam, including three main components: Code Regularization (CR, Section ~\ref{subsec:Regularization}), Neural Translation Model (NT, Section~\ref{subsec:NT Model}), Semantic Information Recovery (SR, Section~\ref{subsec:Semantic Information Recovery}).
In the CR phrase, \fname is committed to \iname generation and Statement Canonicalization for building a regularized dataset that could help NMT models better learn transformation rules between low-level PL and high-level PL.
In the NT phrase, we design a new neural network architecture \mname 
that is suitable for learning transformation rules between low-level PL and high-level PL, which trains based on the regularized dataset to learn the transformation rules accurately. Then we build a classification model named \textit{BinSeg} to separate low-level PL into code snippets, and finally, \fname utilizes \mname to decompile the code snippets sequentially and outputs a C-like high-level PL sketch with the same functionality. 
In the SR phrase, \fname further improves the readability of decompiled high-level PL code through meaningful variable name replacement.

Regularization aims to compensate for the information asymmetry between high-level PL and low-level PL, as mentioned above, which allows \mname to learn the conversion rules accurately. In order to reduce the structure difference between high-level PL and low-level PL, we design a new intermediate language \iname for CR (step \textcircled{1}), which can transform the hierarchical high-level PL into a serialized representation similar to the low-level PL. To make up for the semantic asymmetry between high-level PL and low-level PL, we propose a statement canonicalization approach for CR (step \textcircled{2}), which automatically recognizes identifier types and uniformly replaces identifiers in the PL in units of type for reducing their negative impact on \mname training. Secondly, \mname uses the regularized \iname and low-level PL pair as the training set to learn transformation rules between low-level PL and high-level PL (step \textcircled{3}). 

After the model training is completed, we can use \mname to decompile the real-world low-level PL. In order to avoid that the target low-level PL code is too long and affects the translation effect, we propose a segmentation approach that constructs PL in units of high-level PL statement (step \circled{1}), similar to the translation in natural language use sentence as units. Then we perform statement canonicalization (step \circled{2}) and translate the low-level target PL in terms of segments. Further, a functionally similar high-level PL sketch is obtained (step \circled{3}). As the identifiers in the PL sketch translated by \mname are regularized without semantic information, we design a semantic information recovery approach for SR (step \circled{4}), which enables generating identifier names with semantic information based on the functionality of the PL. After the identifier names are embedded into the high-level PL sketch based on probability priority, we get the decompiled high-level PL with semantic information and similar functionality (step \circled{5}).

\subsection{Code Regularization}
\label{subsec:Regularization}

As mentioned previously, one main challenge of applying the NMT model to decompilation is to compensate for the information asymmetry between low-level PL and high-level PL. In order to overcome this challenge, instead of directly feeding the pair of low-level PL and high-level PL into the model for learning their relationship, we perform code regularization to diminish the impact of the information asymmetry. On the side of high-level PL, we construct an intermediate language called \iname that builds on the AST structural features of the PL, making the high-level PL has a similar expression form as the low-level PL, which explicitly expresses the semantic information in the high-level PL. On the side of low-level PL, the identifiers are regularized for reducing their adverse effects in model training. The PL goes through the above two steps of regularization to make up for the asymmetry of the structure and semantics between low-level PL and high-level PL, allowing the model to learn better the conversion rules between high-level PL and low-level PL.

\vspace{5pt}\noindent\textbf{\iname Generation}.
The central information asymmetry between high-level PL and low-level PL is the semantics implicitly held by high-level PL while explicitly expressed by low-level PL. Such a difference is unique to PL and rarely appears in natural languages (e.g., English and Spanish), which is quite hard to learn by NMT. For example, in the following source code \texttt{a=(a+b)*c+d*e}, the brackets indicate the \texttt{+} operation (i.e., \texttt{(a+b)}) should be executed before the \texttt{*} operation (i.e., \texttt{b*c}). Also, in the expression, the second \texttt{+} operation (i.e., add the result of \texttt{(a+b)*c} and the result of \texttt{d*e}) should be executed in the last step. In contrast, the semantics is directly expressed by the low-level PL. For example, the \texttt{+} operation is directly expressed by add eax,rpb[0x04] before the \texttt{*} operation; and the second \texttt{+} operation is executed in the last step. It is tough for NMT to capture such implicit semantics, which leads to errors in the translated results (e.g., wrong positions of the brackets). So before the translation, we should feed the NMT such implicit semantics to make the results more accurate.

\begin{figure}
	\centering
	\setlength{\abovecaptionskip}{10pt}
	\setlength{\belowcaptionskip}{0pt}
	\includegraphics[width=0.95\columnwidth]{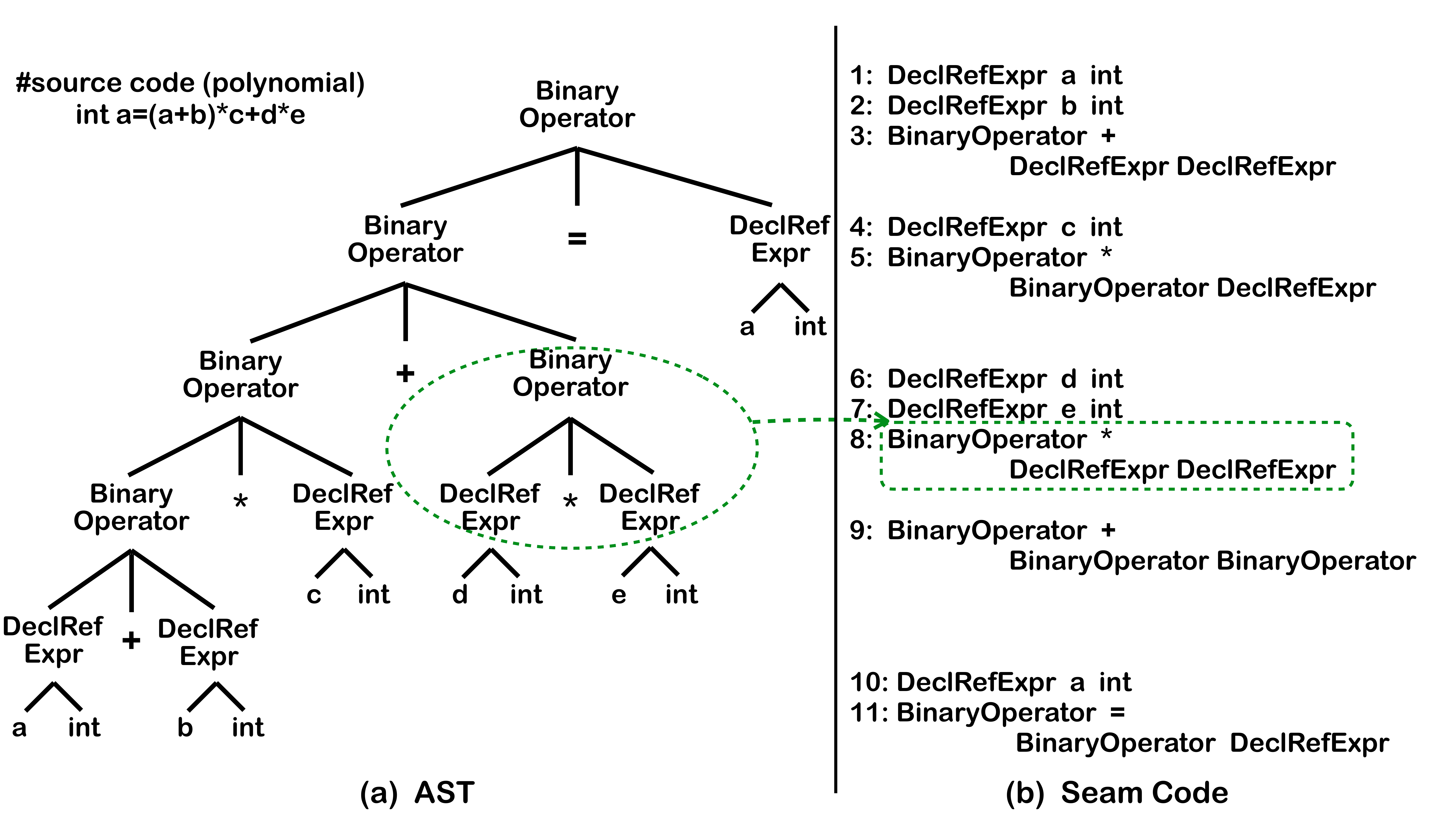}
	\caption{An example of \iname generation}
	\label{pic:IR-generation}
	\vspace{-15pt}
\end{figure}

In order to handle the above problem of information asymmetry caused by implicit semantics, we design an intermediate language called \inam. It should explicitly express the semantics just like the low-level language, from which NMT can benefit from translation. Also, it should be akin to the high-level PL that can be easily transformed back to the high-level PL after the NMT translates a piece of binary code to \inam. For example, the stack operation (e.g., \texttt{push} and \texttt{pop}) and direct memory operation should no longer exist in \inam. Inspired by the design of AST that represents PL syntactic structure in a tree-like form, we propose to generate \iname from AST, making use of AST to express the sequence of operation and the connections between variables. Considering the importance of type information for high-level PL, we keep it in the AST, which gives the NMT the opportunities to learn such high-level semantics. Also, considering that there are too many constants (immediate numbers and strings) in high-level PL that cause NMT to build a huge dictionary, we use a placeholder for constants in AST to decrease the burdens of NMT in the learning process. Similarly, the target branches in \texttt{if}/\texttt{while}/\texttt{for} statements or target functions in function calls are replaced with placeholders. Note that such missing information will be recovered by semantic analysis of the low-level PL (see Section~\ref{subsec:Semantic Information Recovery}).
Figure~\ref{pic:IR-generation} gives an example of \iname generation, (a) is the AST of \texttt{a=(a+b)*c+d*e}, and (b) is the \iname for the left example expression.

\vspace{5pt}\noindent\textbf{Statement canonicalization}. 
Similar to use placeholders in the \inam, we also replace the constants and targets of jump/function call with the placeholders in low-level PL due to their negative impacts on the accuracy of NMT. The immediate numbers are replaced with the placeholder \texttt{IMM}, and constants are replaced with \texttt{STR}. Different from handling the constants, we distinguish the different directions of jump instructions (e.g., \texttt{jmp}, \texttt{jr}, \texttt{je}, etc.). Particularly, we compare the target address of a jump instruction ($J_d$) with the address of the jump instruction itself ($J_i$), if $J_i < J_d$, we replace the target address with the placeholder \texttt{up}. Otherwise, we use the placeholder \texttt{down}. In this way, the NMT is able to distinguish the branch statements (e.g., \texttt{if}) with the loop statements (e.g., \texttt{for}, \texttt{while}). Regarding the function calls, we use the placeholder \texttt{FUNC} for all of the targets.

Based on our evaluation, the results show that the introduction of \iname and statement canonicalization to handle the information asymmetry is sufficient for the accurate translation (i.e., increasing accuracy by 36.10\%).

\vspace {-5pt}
\begin{table}
\centering
\footnotesize
\caption{Statement canonicalization rule}
\label{tab:statement canonicalization}
\begin{tabular}{m{3cm}
<{\centering}|m{3cm}
<{\centering}}
\hline
\textbf{Identifier}& \textbf{Replacement}\\
\hline \hline
\textbf{immediate} & \text{IMM}\\ \hline
\textbf{string} &\text{STR} \\ \hline
\textbf{jump direction} &\text{up/down} \\ \hline
\textbf{function name} &\text{func} \\ \hline

\end{tabular}
\end{table}


\subsection{Neural Translation Model}
\label{subsec:NT Model}

Given a low-level PL code segment $AC$, we generate the corresponding \iname segment $SC$. $AC =\{Insn_1, Insn_2,..., Insn_n \}$, where $ Insn_i$ represents the i-th instruction in the code segment, and $n$ represents the number of instructions contained in the code segment. Each instruction is represented as $Insn_i = \{m, s_1, s_2,..., s_k \}$, where $ m $ represents the instruction mnemonic, and $ s_i $ represents the i-th word in the instruction operand string words, and $k$ represents a total of $k$ words in the operand. $ SC =\{Stmt_1, Stmt_2,..., Stmt_m \} $, $ Stmt_i\in STMT $, where $Stmt_i$ represents \iname instructions, $STMT$ represents the set of all \iname instructions, which is a limited set for a specific instruction. In the high-level PL, $m$ means there are a total of $ m $ instructions in the \iname fragment.

\begin{align}
p(SC) & = p(SC| AC) \\
& =p(<sos>,Stmt_1,...,Stmt_m|AC) \\
& =\prod_{i=1}^m p(Stmt_i|<sos>,...,Stmt_{i-1},AC)
\end{align}

where $<sos>$ represents the start of sentence. 


\vspace{5pt}\noindent\textbf{Model architecture}. A straightforward idea of designing the model is to use a long short-term memory network (LSTM)~\cite{sutskever2014sequence}, which is good at handling strings. However, the main constrain of LSTM is that LSTM does not work well for long sentences (e.g., more than 50 words) and cannot be calculated in parallel. In PL, different words have different effects on other words. The self-attention-based model Transformer~\cite{vaswani2017attention} can capture the weights between words better than LSTM. Furthermore, Transformer allows parallel computing (avoid recursion), which significantly reduces the training time. 

\begin{figure}
	\centering
	\setlength{\abovecaptionskip}{8pt}
	\setlength{\belowcaptionskip}{0pt}
	\includegraphics[width=0.85\columnwidth]{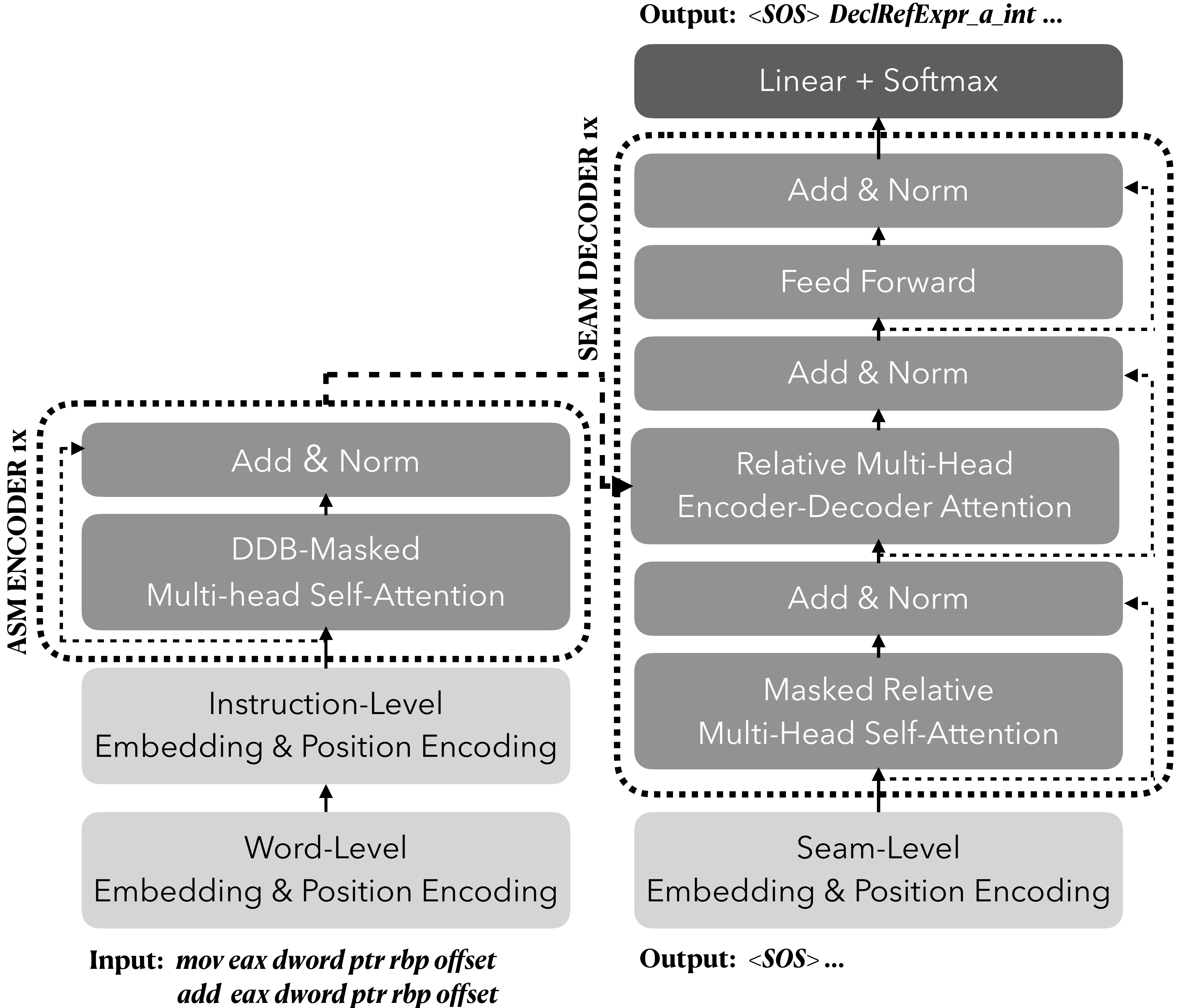}
	\caption{Model architecture}
	\label{pic:model-arch}
	\vspace{-10pt}
\end{figure}




We design a neural decompilation model based on the Transformer, shown in Figure \ref{pic:model-arch}. The model consists of an encoder and a decoder. The encoder extracts features of $AC$ and outputs a feature vector, which committed to decodes $Stmt_i$ of the $SC$ sequence one by one. Then, taking into account the characteristics of the code, we design our embedding and attention scheme.

\begin{figure}
	\centering
	\setlength{\abovecaptionskip}{8pt}
	\setlength{\belowcaptionskip}{0pt}
	\includegraphics[width=0.7\columnwidth]{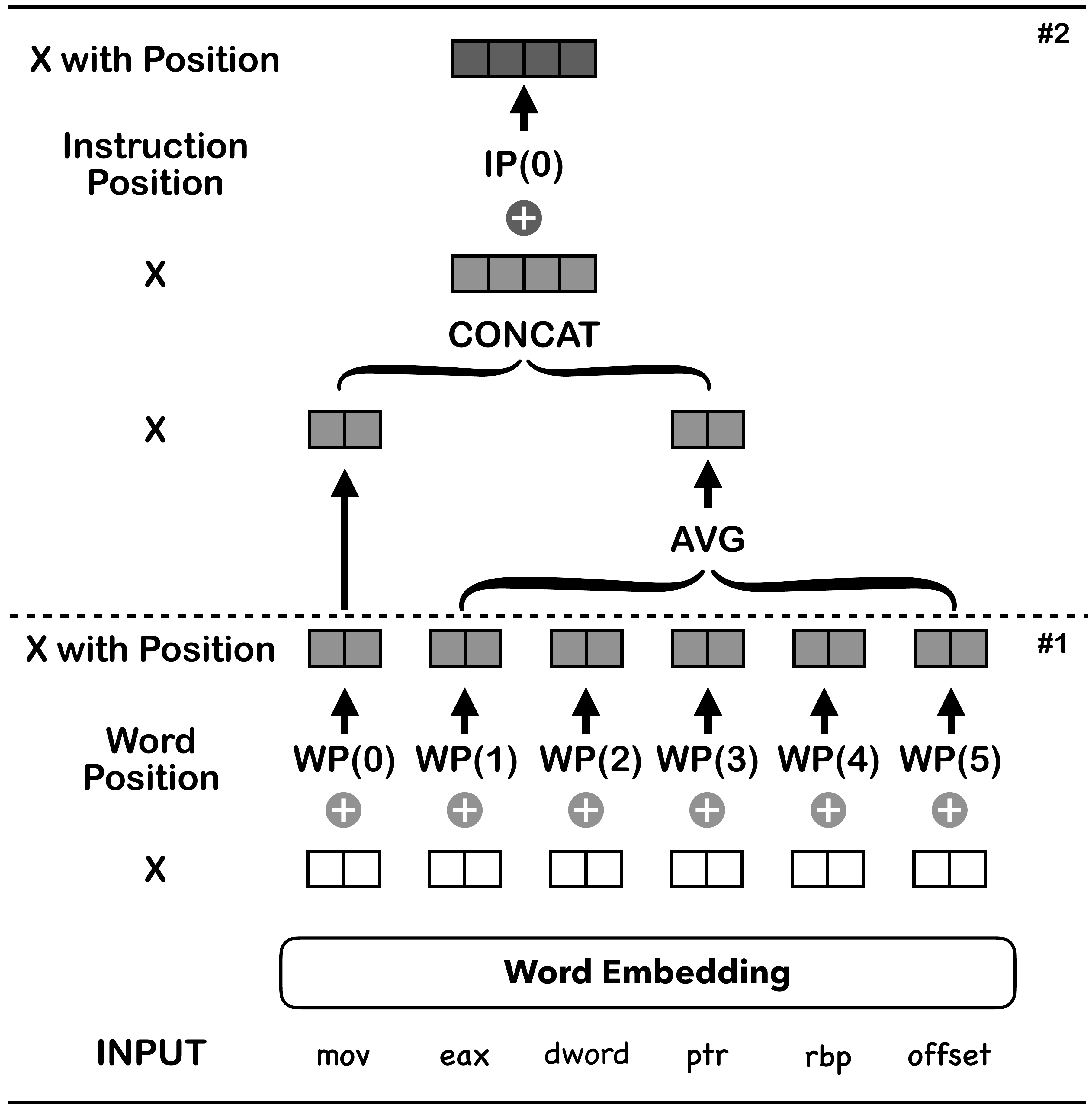}
	\caption{Embedding and position encoding}
	\label{pic:emb}
\end{figure}

\vspace{5pt}\noindent\textbf{Embedding scheme}. 
There is a boundary between two instructions in the ASM. However, the sequence model will put these instructions together during translation. Unfortunately, simple punctuation does not convey this boundary information very well. The mnemonic and operand in instruction are generally highly correlated, and there is usually an overall relationship between the two instructions. The relationship between instructions is complex, as an instruction exists as a whole, and its internal components have their semantics. When embedding an instruction, if we encode an instruction as a whole, the semantics of its internal components will be lost. Conversely, if we are embedding each word within an instruction individually, the overall semantics of the instruction will be lost. The result of Asm2Vec~\cite {ding2019asm2vec} shows that the combined feature vector of the entire instruction can better express the semantics of the code. Therefore, as shown in Figure~\ref{pic:emb}, we design a two-step embedding and position encoding scheme to embed the input based on the above characteristics.

Step 1: For each instruction $Insn_i$ in $AC$, we use the word embedding scheme to get $x_i = \{x_{i0},x_{i1},...,x_{ik}\}$. Where $x_{i0}=emb(m)$, $x_{ij}=emb(s_j)$. Position information inside the instruction indicates the components of the word, such as the mnemonic, source and destination operand, usually have fixed positions. We add absolute position to $x_i$ to get new embedding with positions $x_i=\{x_{i0}+p_0,x_{i1}+p_1,...,x_{ik}+p_k\}$, where $p_j=pos(s_j)$.

Step 2: 
For each $x_i$ obtained in step 1, we calculate the new $x_i=concat(x_{i0}+p_0,Avg_{j=1}^k(x_{ij}+p_j))$. For each instruction $insn_i$ in $AC$, the relative position information between instructions can better indicate the operations' sequence. Therefore, we use relative position-encoding in step 2. Finally, the vector of $AC $ is expressed as $X=\{x_1;x_2;...;x_n\}$.

\begin{figure}
	\centering
	\setlength{\abovecaptionskip}{8pt}
	\setlength{\belowcaptionskip}{0pt}
	\includegraphics[width=0.85\columnwidth]{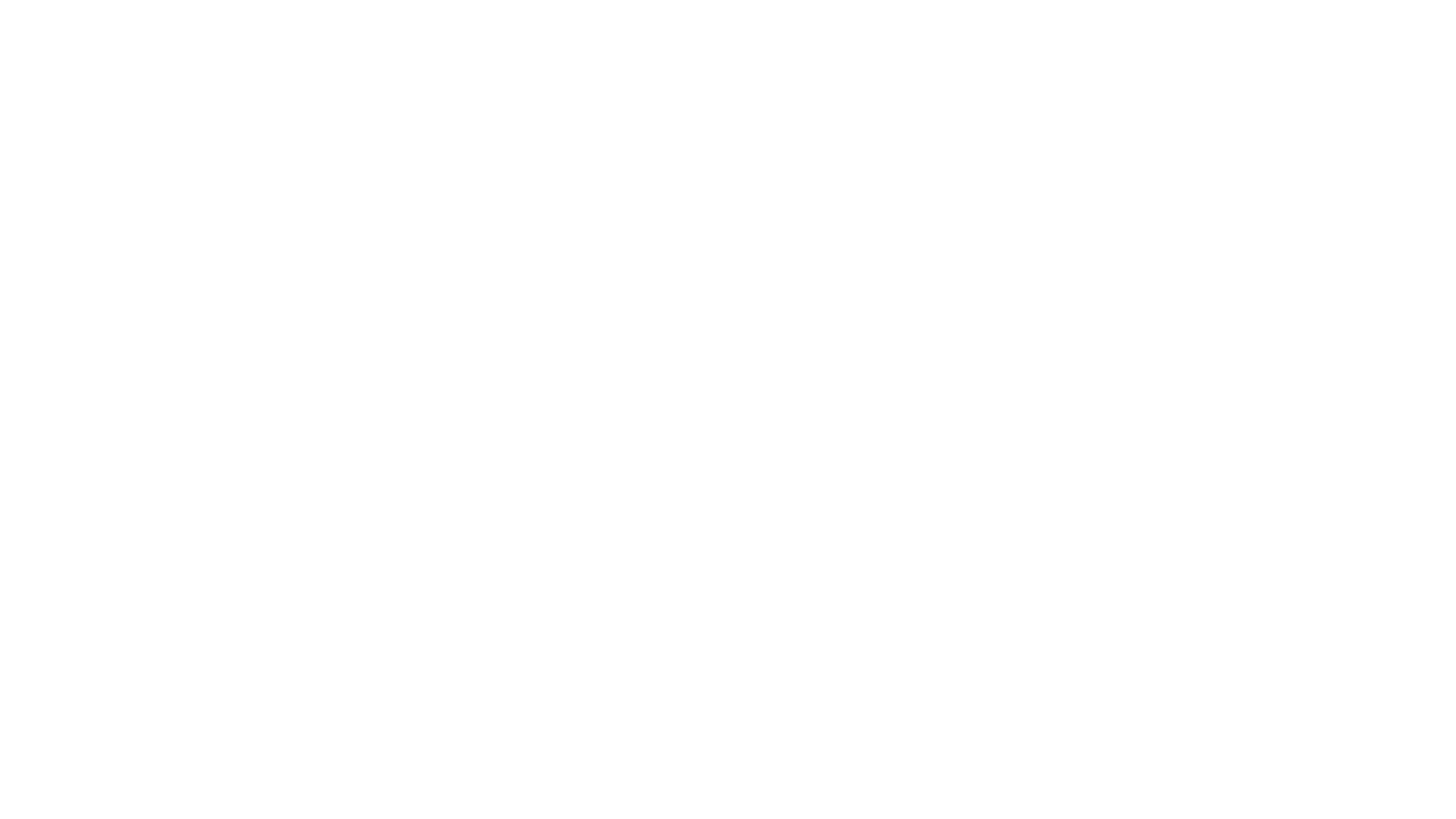}
	\caption{Attention mechanism}
	\label{pic:mask}
	\vspace{-15pt}
\end{figure}

\vspace{5pt}\noindent\textbf{Attention scheme}. 
Unlike natural language, if there are data dependencies between two PL instructions, they are closely related, and vice versa. Therefore, we propose a data-dependent mask to solve this problem. Our attention scheme is shown in Figure \ref{pic:mask}. 

\begin{equation}
{
Attention(Q,K,V)=softmax(\frac{QK^T}{\sqrt{d_k}}\bigodot M)V
}
\end{equation}

where $M$ is a $n * n$ matrix, and 

$$m_{ij}=
\begin{cases}
1, if Related (Insn_i, Insn_j) \\
0, otherwise
\end{cases}
$$

To this end, we use cfile~\cite{cfile} to generate 1,600,000 pairs of inputs and outputs as the dataset for training \fnam. Benefit from the design of \textit{BinTran}, the accuracy of our approach is increased by 44.14\%, compared to the state-of-the-art natural language translator (Google's Transformer~\cite{vaswani2017attention}).

\vspace{5pt}\noindent\textbf{Code segmentation}. 
In the translation (i.e., decompilation) process, to further increase the accuracy, given a piece of assembly code (usually corresponding to many lines of high-level source code), we should separate it before the translation. Each segmented assembly code should be exactly corresponding to one line of source code in the ideal situation. This is similar to the translation of natural language. If a paragraph is not separated into sentences, it is pretty hard for translation, even not easy for human readers to understand. So just like separating natural languages using punctuation (e.g., period, comma, etc.), we separate low-level PL into code snippets.

The basic idea is based on the observation: the ending statement of the assembly code segment that corresponds to each line in the high-level PL has certain features connected with the previous instructions. Note that handling the inputs is similar to the process of translating low-level PL. So we build a classification model called \textit{BinSeg} based on only the encoder part of \textit{BinTran}, which takes assembly code as input by labeling each line of low-level PL to determine whether it is a boundary. Particular, \textit{BinSeg} contains 1 layers. Then we still use the dataset, including 1,600,000 code snippets as the dataset for training. \textit{BinSeg} is demonstrated to be very accurate, with an accurate rate of 98.64\%.

\subsection{Semantic Information Recovery}
\label{subsec:Semantic Information Recovery}

\begin{figure}
	\centering
	\setlength{\abovecaptionskip}{10pt}
	\setlength{\belowcaptionskip}{0pt}
	\includegraphics[width=0.85\columnwidth]{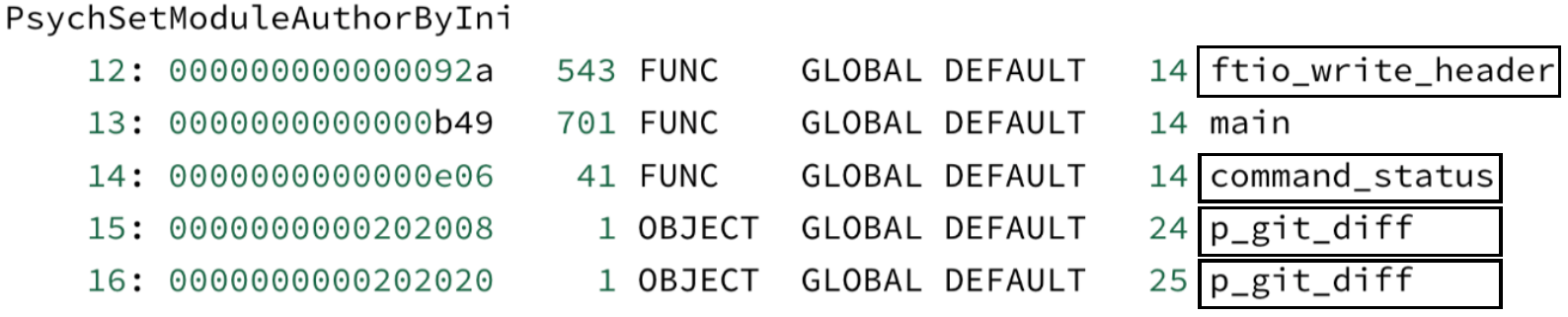}
	\caption{An example of using DEBIN to recover variable names}
	\label{figs:debin-example}
	\vspace{-10pt}
\end{figure}

After the decompilation of the target low-level PL by \textit{BinTran}, we obtain a high-level PL sketch. Recall that we use \iname in the training process. So the output of \textit{BinTran} is not the final high-level PL, which only functionally corresponds to the target low-level PL, but without the semantic information of identifiers such as function names, variable names. One may directly analyze the binary code and use arbitrary variables in the high-level PL, which may significantly impact the readability, just like the source code translated from Hex-Ray~\cite{Hex-Rays}.
To improve the readability of the decompiled high-level PL, several approaches are proposed for recovering the semantic information of decompiled high-level PL~\cite{jaffe2018meaningful, he2018debin,lacomis2019dire}, which aims to recover information such as variable names. However, they are not effectively integrated with decompilers. The binary debugs information recovery approach~\cite{he2018debin} can recover identifier semantics by making predictions from an existing set of identifier names, which has significant limitations and relies on information such as the symbol table in the current binary file. We conducted a practical test on DEBIN~\cite{Debin} and found that its recovery effect is not ideal, as shown in Figure \ref{figs:debin-example}. The low-level PL we tested implements the Fast Number Theory Transformation (FNTT) algorithm, which contains operations such as Discrete Fourier Transform (DFT) and multiplication. The corresponded high-level PL contains four functions, \textit{dft}, \textit{pre}, \textit{qpow}, and \textit{main}. However, the name of the DEBIN prediction (shown in the figure) is far from the function of the tested PL. Accurately recovering such semantic information of identifiers is by no means trivial. 

\begin{figure}
	\centering
	\setlength{\abovecaptionskip}{8pt}
	\setlength{\belowcaptionskip}{0pt}
	\includegraphics[width=1.0\columnwidth]{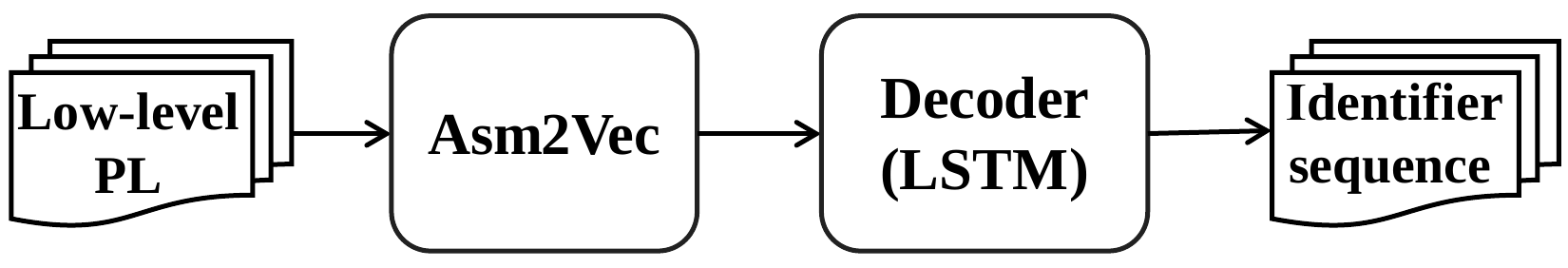}
	\caption{The process of identifier recovery}
	\label{pic:identifier-renaming}
	\vspace{-10pt}
\end{figure}

\vspace {5pt}\noindent\textbf{Semantic-based identifier recovery.}
To achieve accurate recovery of the semantic information of the decompiled high-level PL, we propose a semantic-based approach to recover identifier names that rely solely on the low-level PL itself and require no other additional information. 
Drawing on the idea of image caption generation~\cite{vinyals2015show} in NLP, we define the problem of recovering identifiers in decompiled PL functions as an encode-decode problem and build an encoder-decoder model for automated generation of identifier sequences $S$. Different from \textit{BinTran} and \textit{BinSeg} that need to take attention to different regions of the inputs, the model here only needs to abstract the input low-level PL. Therefore, in the design of this model, we leverage LSTM~\cite{LSTM}, which is suitable for decode sequence. In particular, we utilize Asm2Vec~\cite{ding2019asm2vec} as the encoder, which is responsible for encoding the input low-level PL into fixed-length (d=2048) vectors in units of function. Further, we use LSTM~\cite{LSTM} as the decoder, which aims to decode the vectors to generate the corresponding named sequences. The structure of the semantic information recovery model is shown in Figure \ref{pic:identifier-renaming}. The reason for using Asm2Vec is that it allows a function to be expressed as a vector, and the vectors can represent the structure and semantic information of the function. For functions with similar functionalities, Asm2Vec can generate relatively similar vectors, which coincides with the programming specifications. 

The model takes low-level PL function $F_{asm}$ as input and outputs an identifier sequence $S$. Based on the methods above, training is first performed using \textit{(S, $F_{asm}$)} and then the decoding task is completed using the LSTM model.

\begin{figure}
	\centering
	\setlength{\abovecaptionskip}{10pt}
	\setlength{\belowcaptionskip}{0pt}
	\includegraphics[width=0.8\columnwidth]{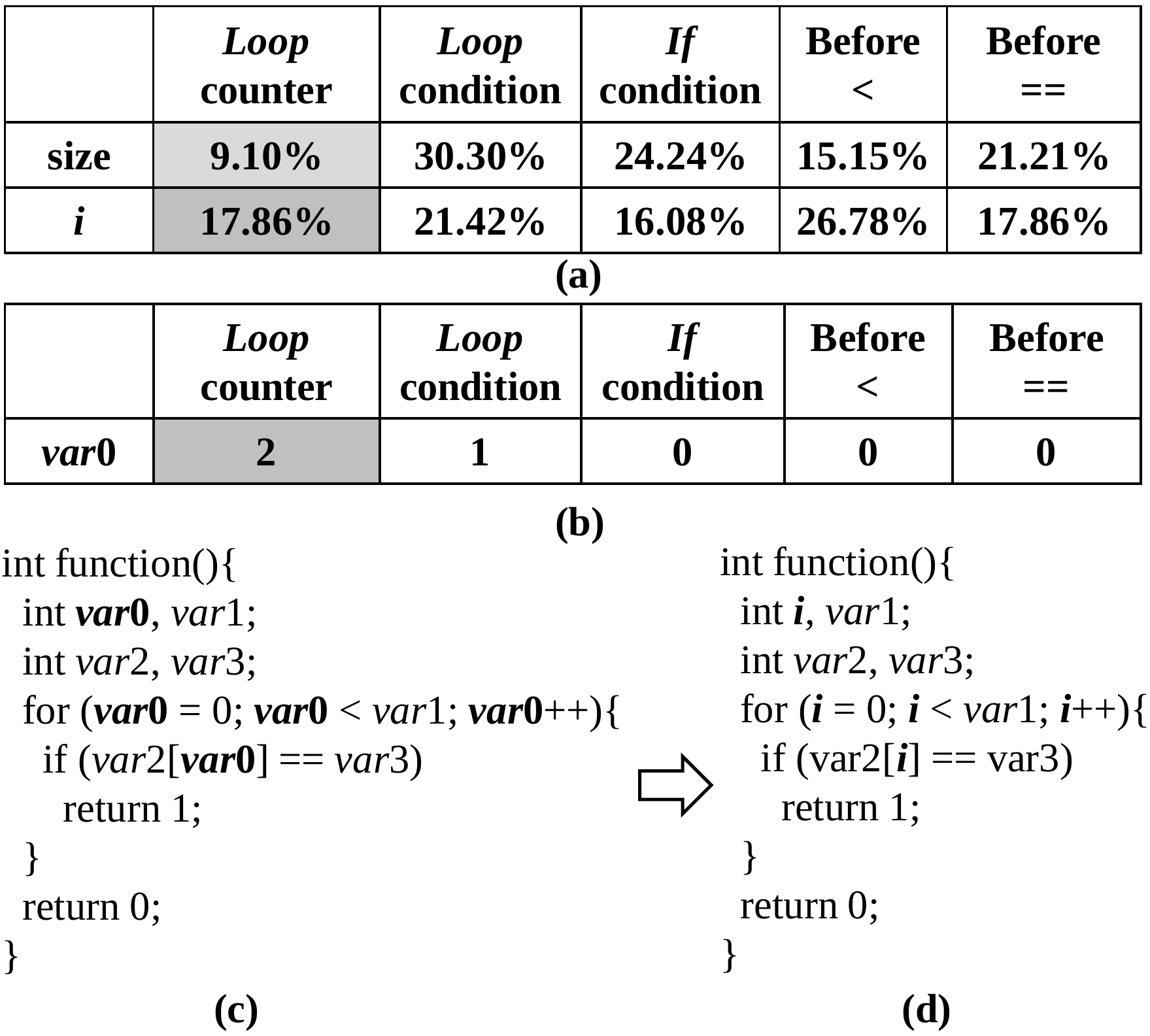}
	\caption{An example of semantic identifier embedding}
	\label{figs:embed-example}
	\vspace{-10pt}
\end{figure}

\vspace {3pt}\noindent\textbf{Semantic identifier embedding.}
After generating meaningful identifiers, we embed them into the translated high-level PL. We assign the identifier to appropriate locations according to the probability priority learned from our previous dataset.
Particularly, for each function, we check 26 common positions where variables are usually placed. Those positions include 5 types, (1) \texttt{Loop} counter, (2) \texttt{Loop} condition, (3) \texttt{If} condition, (4) before or after \texttt{[+, -, *, /, \%, >, <, >= <=, ==, !=]}, (5) \texttt{return}.

Our method can divided into 6 steps: (1) Count each identifier in the training set used in the identifier sequence generator model and the number of its occurrences at the above positions in functions, and obtain a dictionary $vocab: \{id_1: [p_0,...,p_{25}], ..., id_n: [p_0,...,p_{25}]\}$, $p_i$ means frequency that $id_j$ appeared in position \texttt{i}. (2) Count the number of times each identifier appears in the above position in translated function, and get a dictionary $function: \{var_0: [p_0, ..., p_{25}], ..., var_m: [p_0, ..., p_{25}]\}$. (3) After get the model generated identifier sequence S, We can get a subset of $vocab$, named $vb$, where $vb$ contains only the items corresponding to the identifiers appearing in the sequence $S$ in vocab. (4) For every $var_i$ in function, we can get a max possible location $pos$. (5) For every variable in $function$, we choose the identifier with the highest probability corresponding to the $pos$ in $vb$ as the variable's name. (6) If it is not found, we choose from the position with the next lowest probability and repeat the above operation until the variable is assigned an identifier. If the candidate identifier generated by the model is insufficient, the regularized names [$var_1$, ...,$var_n$] are used instead. 

Figure~\ref{figs:embed-example} shows an example of assigning identifiers using the method above. In the figure, (a) shows 2 identifiers \texttt{size} and \texttt{i} generated by the model and the probabilities of their appearance in 5 positions. (b) shows \texttt{$var_0$}, a variable of function showed in (c), and frequency it appears in $5$ positions. As shown in (b), \texttt{$var_0$} appears most frequently in the \texttt{loop} variable position in this function. Therefore, the identifier \texttt{i} with the highest probability of appearing at this location is assigned to \texttt{$var_0$}. For other positions in the function, \fname performs similar operations on other locations to complete the entire function's semantic recovery.



%% file: Implementation.tex
\section{Implementation}
\label{sec:Implementation}

We implemented our technique, which takes the binary executable as inputs and further uses the disassembly tool angr~\cite{shoshitaishvili2016state} for the x86\_64 architecture to generate the corresponding assembly code, which in turn feeds it into \fnam. A high-level PL with semantic information can be generated by sequentially executing the above three modules, and its function is similar to the input low-level PL. Specifically, C is in our implementation of high-level PL.

\vspace{5pt}\noindent\textbf{Code regularization}. The code regularization is responsible for regularizing the dataset used for \mname training. The purpose of code regularization is to make up for information asymmetry between low-level PL and high-level PL so that \mname can better learn the transformation rules. The first step is to reduce the structural asymmetry problem between the high-level PL and the low-level PL by transforming the hierarchical high-level PL into a serialized representation similar to low-level PL. We use \texttt{clang}\footnote{https://clang.llvm.org/}10.0.0 to generate the corresponding AST for the high-level PL, and further, construct an \iname generation tool which converts the AST of high-level PL into an intermediate language \iname with a relatively similar assembly structure. Therefore, the structural asymmetry problem between low-level PL and high-level PL can be reduced. We generated the corresponding AST at the same time as angr disassembled the binary executable. These two modules are independent of each other, and we can speed up the regularization by processing the two parts in parallel. To make up for the asymmetry in identifier semantics between the low-level PL and high-level PL and avoid the influence of identifiers on the effectiveness of \mname training, we construct a regular matching-based tool for the identification and substitution of identifiers, and the rule of statement canonicalization is shown in Table \ref{tab:statement canonicalization}.

\vspace{5pt}\noindent\textbf{Neural translation model}. The \mname Translation involves model initialization and model implementation. In the initial stage, \fname starts to train \mname only after sufficient data is collected or generated. We randomly generate enough training data using cfile~\cite{cfile} and ensure an even distribution of the various types (see Section~\ref{subsec:Experimental Setup}), which is a C code generator implemented with python. Further, we implement the regularization of the training set through the operations of \iname Generation and Statement Canonicalization that in Code Regularization module, aims to reduce the difficulty of \mname learning transformation rules. In the implementation stage, we implement an attention-based model \mname with the help of TensorFlow~\cite{tensorflow2015-whitepaper}. It contains an encoder and a decoder, each of which is composed of a single layer, the specific composition within the layer as mentioned in Section \ref{subsec:NT Model}. In the actual low-level PL translation, in order to avoid the accuracy of the translation being compromised by the length of the function code, we use the encoding module of the Transformer~\cite{vaswani2017attention} to implement a low-level PL partitioning model, which can use the high-level PL line as a unit to achieve the segmentation of the assembly code. We selected 100,000 assembly instructions and labeled the data according to the information provided by the GCC compiler debugging options, then use \texttt{<instruction, label>} as input for model training.

\vspace{5pt}\noindent\textbf{Semantic information recovery}. To achieve semantic information recovery of identifiers, we construct an identifier recovery model drawing on the idea of the image caption generation~\cite{vinyals2015show}. The model is an encoder-decoder architecture. We use Asm2Vec as the encoder and LSTM as the decoder. Asm2Vec encodes the input low-level language into a vector in function units, and LSTM further decodes the vector to predict the identifier sequence corresponding to the function. We collected enough data in different Github algorithm code repositories containing 20,000 functions and using them as the training set. For the immediate count, we fill it in by comparing it with the original assembly statement. 


%% file: Evaluation.tex
\section{Evaluation}
\label{sec:Evaluation}

\subsection{Experimental Setup}
\label{subsec:Experimental Setup}

We evaluated the performance of \fname on a variety of benchmarks with real-world applications and different complexity levels, as shown in Table~\ref{lab:Results on Real-world Projects} and  Table~\ref{lab:Performance under different levels of complexity}. All the experiments are performed on a 64-bit server running Ubuntu 18.04 with 16 cores (Intel(R) Xeon(R) CPU E5-2620 v4 @ 2.10GHz), 128GB memory, 2TB hard drive and 5 GPUs (2 GTX Titan-V GPU and 3 GTX Titan-X GPU).

\vspace{5pt}\noindent\textbf{Training data generation}. To build dataset for training, validation and testing of \fnam. we randomly generate 1,600,000 pairs of high-level PL with corresponding assembly codes. The program is compiled using \textit{GCC 6.5.0} without optimization. The generated high-level PL are arithmetic expressions, logical expressions, \texttt{if} condition, \texttt{while} loop, and \texttt{call} type. There are three levels of complexity for each type: $level\ 0$: var and var, $level\ 1$: var and exp (ex. var/(var + var)), and $level\ 2$: exp and exp (ex. (var+var)/(var+var)). For \texttt{if} and \texttt{while} type, the complexity refers to the complexity of their conditional statements. The maximum depth of the AST representation for each statement in the high-level PL is \cnum{5}. To verify the effect of dataset on the accuracy of \mname translation, we choose the statement size of the training set as 100,000, 200,000, 400,000, 800,000, and 1.6 million. We also ensure that there is no overlap between the test set and the training set.

\vspace{5pt}\noindent\textbf{Benchmarks}. 
We evaluate the performance of \fname using real-world applications. Notably, we select 4 real-world projects: (1) Hacker’s Delight loop-free programs~\cite{warren2013hacker} that is constructed by Schkufza et al.~\cite{schkufza2013stochastic}, which is used to encode complex algorithms as small, loop-free sequences of bit-operated instructions. (2) Libpcap~\cite{Libpcap}, which is a portable C/C++ library for network traffic capture on Linux. (3) GMP~\cite{GMP}, an abbreviation for the multi-precision arithmetic library, a library for arithmetic operations. (4) The solution set of LeetCode~\cite{LeetCode-in-pure-C}.

To further evaluate the performance of \fname on different operations, we use 4 tasks of arithmetic and logical expressions, \texttt{if} condition, \texttt{while} loop, and \texttt{call} type. Each task is evaluated separately from 3 levels of complexity. We choose to evaluate \fname from these four tasks because they are the building blocks of high-level PL, and the ability of \fname to decompile low-level PL code can be objectively reflected through the combined performance of the four tasks.

\vspace{5pt}\noindent\textbf{Evaluation metrics}. We evaluate \fname using two level metrics: (1) word accuracy and (2) sequence accuracy. Word accuracy is the percentage of the predicted words in the high-level PL that match the ground-truth words. Sequence accuracy indicates the proportion of correctly translated sequence to the entire sequence. Since the output of the \mname is a high-level PL code sketch, we ignore the identifier component of the performance evaluation, which will be evaluated in Section~\ref{subsec:effectiveness}.

\subsection{Effectiveness}
\label{subsec:effectiveness}

\vspace{5pt}\noindent\textbf{Performance on real-world projects}. We evaluate \fname using the 4 real-word applications mentioned in Section \ref{subsec:Experimental Setup}. To ensure the accuracy and objectivity of evaluation results, we remove duplicates data. From Table \ref{lab:Results on Real-world Projects}, we can find that \fname is very accurate (ranging from $91.07\%$ to $100\%$). The incorrect rate for all codes is $5.59\%$ on average. Compared with LSTM, our approach achieves 67.57\% higher accuracy on average, mainly because LSTM cannot handle long-term information. \fname achieves over 80\% higher accuracy. 

\vspace{5pt}\noindent\textbf{Performance under different levels of complexity}. We evaluate the performance of \fname on a benchmark of \cnum{4} types of expressions with different complexity levels. As can be seen from the results in Table \ref{lab:Performance under different levels of complexity}, \fname achieves high accuracy in all benchmark tests, even for various types of high complexity expressions ($level\ 3$).
We manually analyzed the sentences that is not correctly translated in the two experiments above. The main reason for the errors of these sentences is that the comparators and function parameters are not translated correctly. Since the comparison operations in the low-level PL code are implemented through the relevant instructions of \texttt{jmp} (e.g.,. \texttt{jg}, \texttt{jl}), and the instructions which uses does not exactly match the comparison symbols in the source code, so it is difficult for \mname to learn this translation rule. For example, the expression \texttt{a < b} corresponds to \texttt{mov eax, DWORD PTR [rbp-4]; cmp eax, DWORD PTR [rbp-8]; jge .L2} in low-level PL code. In this case, \fname would translate the low-level PL code to the expression \texttt{a >= b}, resulting in an  error of the statements. For function parameters, the parameter list is passed by registers in low-level PL code. However, it is difficult to identify the parameters for \mnam based on registers. We provide a method for user interaction, which can provide translation information by the model for user reference to complete the parameter list.

\begin{table}
\centering
\footnotesize
\caption{Results on real-world projects}
\label{lab:Results on Real-world Projects}
\begin{tabular}{m{2.2cm}
<{\centering}|m{1cm}
<{\centering}|m{1cm}
<{\centering}|m{0.8cm}
<{\centering}|m{1.6cm}
<{\centering}}
\hline
\textbf{sequence level}& \textbf{Hacker’s Delight}& \textbf{Libpcap}& \textbf{GMP}& \textbf{Leetcode solution set}\\
\hline \hline
\textbf{SEAM} & \text{100$\%$}& \text{91.07$\%$}& \text{94.59$\%$}& \text{92$\%$} \\ \hline
\textbf{LSTM} & \text{20.75$\%$}& \text{25.00$\%$}& \text{28.2$\%$}& \text{34$\%$} \\ \hline
\textbf{Transformer} & \text{7.2$\%$}& \text{8.02$\%$}& \text{4.72$\%$}& \text{0$\%$} \\ \hline
\end{tabular}
\end{table}

\begin{table}
\centering
\footnotesize
\caption{Performance under different levels of complexity}
\label{lab:Performance under different levels of complexity}
\begin{tabular}{m{1.1cm}
<{\centering}|m{1.3cm}
<{\centering}|m{1.6cm}
<{\centering}|m{1.4cm}
<{\centering}|m{1.2cm}
<{\centering}}
\hline
\textbf{ }& \textbf{expression}& \textbf{$if$ condition}& \textbf{$while$ loop}& \textbf{$call$ type}\\
\hline \hline
\textbf{$Level\ 0$} & \text{99.71$\%$}& \text{99.2$\%$}& \text{96.21$\%$}& \text{-}\\ \hline
\textbf{$Level\ 1$}& \text{93.6$\%$} & \text{96.92$\%$}& \text{96.59$\%$}& \text{-} \\ \hline
\textbf{$Level\ 2$}& \text{90.59$\%$} &\text{88.06$\%$}& \text{91.43$\%$}& \text{74.53$\%$} \\ \hline
\end{tabular}
\vspace{-10pt}
\end{table}

\vspace{5pt}\noindent\textbf{Performance of semantic information recovery.} We evaluate the semantic accuracy of the identifier semantics after identifiers renaming by \fnam. The test set we choose is real-world application Algorithm repository~\cite{Algorithms/C}. We choose the Algorithm repository because the algorithm's functionality is obvious, so the accuracy of identifier semantics is relatively easy to judge. \fname achieves \cnum{92.64}\% accuracy in semantic recovery against identifiers in this real-world project. 

We also evaluate the effectiveness of semantic information recovery through a questionnaire survey. We utilize 7 functions that come from Github algorithm repositories~\cite{Algorithms/C}, using IDA, RetDec and \fname translated code as samples, and invite \cnum{20} technicians from Google, Amazon, Alibaba, and Tencent, as well as \cnum{20} Ph.D. students from our research lab to judge the accuracy and readability. As a result, \fname received a \cnum{99.52}\% praise for readability and \cnum{90.77}\% praise for accuracy. IDA received \cnum{27.14}\% praise for readability and \cnum{94.47}\% praise for accuracy. RetDec received \cnum{7.62}\% praise for readability and \cnum{95.62}\% praise for accuracy. The results show that the semantic information recovered by \fname can help analysts understand the code greatly.

\subsection{Runtime Performance}

We calculate the training time for \mname using the \cnum{1.6} million dataset. The time spent by the model for every 100 steps is 47s, and the model completes the training task with a total of \cnum{20,000} steps. So the training time for \mname is about \cnum{2.6} hours.
Regarding the time of translation, we randomly selected \cnum{100} low-level PL code fragments (corresponding to a line of high-level PL code) from the test set for the translation efficiency test. After \cnum{100} random experiments, the average translation efficiency is obtained. The average time consumed by \mname to translate a line of high-level PL is \cnum{0.32s}. The length of the selected code fragments ranges from 8 to 39, with an average length of \cnum{18}.

\subsection{Influence of Parameters}


\vspace{5pt}\noindent\textbf{Impact under different training set size}. To understand the effect of training dataset size on the performance of \fname and to grasp the demand of \fname on the training dataset, we train the model with different scales respectively. The results are shown in Table \ref{lab:Impact under different training set sizes}, The test dataset used in the experiment is randomly generated. We find that the model already performs well at the 800,000 scale of the dataset, but the performance of the model does not improve significantly as the dataset increases. This informs the scale of the training dataset when the model is migrated to a new PL translation task.

We further compared the effectiveness of two translation methods, Line2Line(src) and Line2Line(AST). The former directly uses the low-level PL and high-level PL pairs as the training data set, while the latter serializes the source code and uses the low-level PL and \textit{SeamCode} pairs as the training data set. The results in Table~\ref{lab:Impact under different training set sizes} show that the latter is on average 35.25\% more accurate than the former, proving that serialized expression of the high-level PL is more conducive to \mname learning (see Section~\ref{subsec:Regularization}).

\begin{table}
\centering
\footnotesize
\caption{Impact under different training set sizes}
\label{lab:Impact under different training set sizes}
\begin{tabular}{m{2cm}
<{\centering}|m{2.4cm}
<{\centering}|m{2.4cm}
<{\centering}}
\hline
\textbf{ }&  \textbf{Line2Line(src)}& \textbf{Line2Line(AST)}\\
\hline \hline
\textbf{10w} & \text{54.50$\%$}& \text{97.64$\%$}\\ \hline
\textbf{20w}& \text{62.54$\%$} & \text{97.79$\%$} \\ \hline
\textbf{40w}& \text{60.08$\%$} &\text{97.93$\%$} \\ \hline
\textbf{80w}& \text{50.80$\%$} &\text{98.16$\%$} \\ \hline
\textbf{160w}& \text{47.60$\%$} &\text{97.75$\%$} \\ \hline
\end{tabular}
\vspace{-10pt}
\end{table}

\vspace{5pt}\noindent\textbf{Impact of position encoding}. 
We evaluate the accuracy of using different position encoding methods, including absolute position encoding proposed by Vaswani et al.~\cite{vaswani2017attention} and relative position encoding proposed by Shaw et al.~\cite{shaw2018self}. The highest accuracy using relative position is~\cnum{98.64}\%, while accuracy using absolute position encoding is~\cnum{96.48}\%. The results show that the accuracy using relative position is higher than using absolute position. In relative position encoding, the value of the max relative position will have a certain effect on the accuracy of the model. We evaluate 4 different values, \cnum{15, 20, 25, and 30} in relative position encoding. We find that the \mname model works best when the maximum relative position is set to \cnum{20}.

\begin{figure}
	\centering
	\setlength{\abovecaptionskip}{8pt}
	\setlength{\belowcaptionskip}{0pt}
	\includegraphics[width=1\columnwidth]{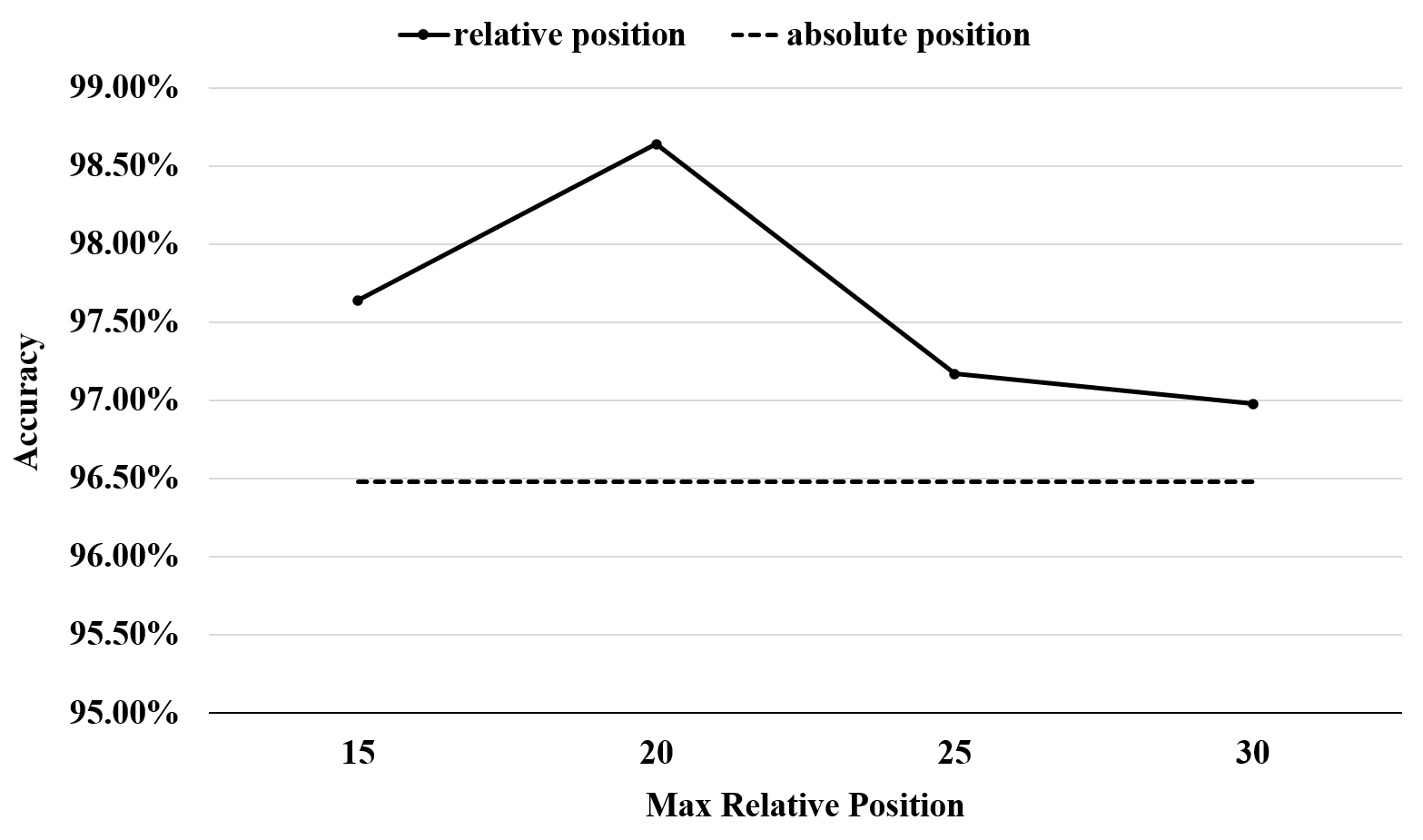}
	\caption{Accuracy of relative position and absolute position}
	\label{figs:position}
\end{figure}

\subsection{Understanding}

Our experiments show that \fname achieves high performance and accuracy, with word accuracy of \cnum{98.65}\%. In addition, to understand whether the semantic representations of the input code are reasonable and interpretable, we analyze the visualization of attention scores to manually check whether the module of multi-head attention focuses on the critical features of the input. The attention score of each location reflects its contribution to the output result. The higher the score, the greater the contribution is. Based on this idea, we observe the model's semantic understanding by visualizing the attention score.

\ignore{In Figure~\ref{}, the darker lines in the figures correspond to higher attention scores.} 

\begin{figure}
	\centering
	\setlength{\abovecaptionskip}{8pt}
	\setlength{\belowcaptionskip}{0pt}
	\includegraphics[width=0.8\columnwidth]{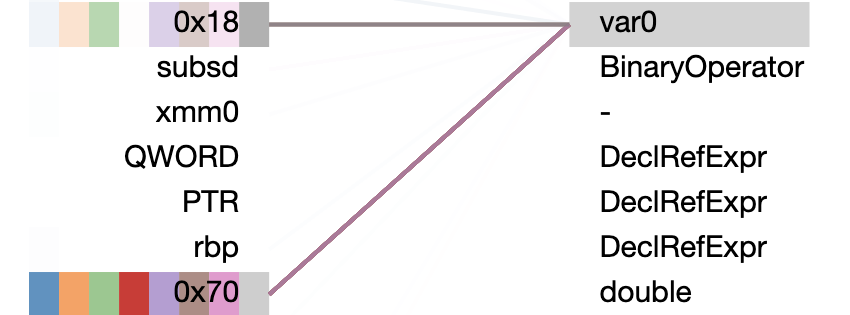}
	\caption{Visualization of attention mechanism}
	\label{figs:understanding}
	\vspace{-10pt}
\end{figure}

As we all know, each variable has its corresponding storage unit in the memory. Different types of variables occupy different sizes of a storage cell. The name of a variable maps to its address which is the lowest address of the storage cell occupied by the variable in memory. In our evaluation, we find that the multi-head attention of the first layer correctly maps the memory address to its corresponding variable name. Figure~\ref{figs:understanding} shows a concrete example. When translating the current position var0 in the figure, the most significant contribution is the memory address 0x70, which makes sense. In the low-level PL code, different define directives reflect different storage spaces reserving for the variables. 

From the above visualization analysis, we can gain an insight into the multi-head attention mechanism and answer whether it pays attention to critical features when decompiling a target low-level PL code. Furthermore, the results sufficiently prove that the semantic understanding learned by our model is reasonable and interpretable.

%% file: Discussion.tex
\section{Discussion}
\label{sec:Discussion}

\vspace{5pt}\noindent\textbf{Limitations}. In our work, we propose and implement a novel attention-based neural framework named \fnam. The evaluation shows that the approach performs well. However, there still exist some limitations. \fname builds independent models for high-level PL sketch translation and semantic information recovery. There is no fusion of the two core engines. However, this independent design also allows the advantage that those two engines could be used separately to assist other tools in translating programs or recovering semantics. For example, the semantic information recovery engine can be applied to tools like Hex-Rays to help recover semantic information for decompiled code. Besides, \fname has poor translation performance for compiler-optimized PL code, which is also powerless for all the-start-of-art techniques. We choose the code from \cnum{50} algorithms from the LeetCode solution of \textit{C} code~\cite{LeetCode-in-pure-C} dataset as the test set and evaluate the performance of \fname against the low-level PL code that is generated from GCC compiler at different optimization levels (\textit{O1} to \textit{O3}). However, we get~\cnum{20}\% accuracy on these three datasets. \fname can translate the optimized code to some extent because it learns several necessary data types to make up the PL.

\vspace{5pt}\noindent\textbf{Future Work}. We will continue to explore techniques for improving the translation effect and semantic recovery accuracy of \fnam, and resolve the above limitations to expand the translation capabilities of \fnam. For example, we will add optimized code data to the training dataset and try to let \mname learn more optimization rules so that \fname can accurately translate more optimized low-level PL codes. Furthermore, we will integrate low-level PL code translation and semantic information recovery into the same model, thereby effectively reducing the complexity of \fnam.

%% file: Related_work.tex
\section{Related work}
\label{sec:Related work}

\vspace{5pt}\noindent\textbf{Conventional decompilation method}. Decompilation originated from the need for software porting. The relevant research can be traced back to the 1960s, the decompiler D-Neliac~\cite{D-Neliac} built by the US Naval Electronics Laboratory for the Neliac program language~\cite{10.5555/1096930}. In the following half-century of research and development, many decompilers were developed, such as Hex-Rays~\cite{Hex-Rays}, Phoenix~\cite{brumley2013native}, RetDec~\cite{kvroustek2017retdec}, and Ghidra~\cite{Ghidra}. Hex-Rays is currently the most mature commercial decompiler and is still considered the de-facto industry standard in the software security industry. RetDec is a redirectable decompiler based on the LLVM architecture developed by the Czech security company Avast, aiming to be the first ``universal'' decompiler to support multiple architectures and program languages. Ghidra is developed by the National Bureau of Security Research, which supports multiple processor instruction sets and executable formats. 

Other works, such as TIE~\cite{lee2011tie}, DIRE~\cite{lacomis2019dire}, DREAM~\cite{yakdan2015no},  DREAM++~\cite{yakdan2016helping}, target at improving the readability of decompiled code by refactoring variable types~\cite{lee2011tie} or names~\cite{lacomis2019dire}, as well as eliminating \textit{goto} statements~\cite{yakdan2015no, yakdan2016helping}.
Although these works have made significant improvements to decompilation, they all need to detect known control flow structures based on manually written rules and patterns, and these rules are difficult to develop, error-prone, usually only capture part of the known CFG, and require long development cycles. Different from these traditional rule-based decompilers, the goal of \fname is based on deep neural networks to automatically learn and extract rules from code data, which is learned from the principles of NMT, trying to break through the problems of poor scalability, complex development, and long cycle of those tools due to over-reliance on manual rule definition.

\vspace{5pt}\noindent\textbf{Neural networks for program analysis}.
The artificial intelligence (AI) technology represented by deep learning has been widely used in program analysis such as function recognition~\cite{rosenblum2008learning, karampatziakis2010static, bao2014byteweight, shin2015recognizing, chua2017neural} and identifier renaming~\cite{jaffe2018meaningful, he2018debin,lacomis2019dire}, improving the degree of automation and intelligence effectively of the software development or analysis process. 
Function identification is the cornerstone of reverse engineering and binary program analysis. The premise of decompilers to provide helpful output is to know the location of the function~\cite{bao2014byteweight}. Rosenblum et al.~\cite{rosenblum2008learning} first converted the function recognition problem in binary programs into a supervised machine learning classification problem. Bao et al.~\cite{bao2014byteweight} proposed a machine-learning and weighted prefix tree-based automatic function recognition system BYTEWEIGHT. Shin et al.~\cite{shin2015recognizing} introduced RNN based on BYTEWEIGHT, proved that RNNs have a specific advantage in binary function recognition. 
Identifier renaming aims to improve the readability of decompiled code. Jaffe et al.~\cite{jaffe2018meaningful} designed a lexical n-gram-based NMT model to automatically assign meaningful names to variables, which is based on vocabulary information obtained by tokenized code. DIRE~\cite{lacomis2019dire} further integrates the structural information of the AST corresponding to the tokenized code.
Different from their work, we recovery semantic information based on functional characteristics of functions.

\vspace{5pt}\noindent\textbf{Neural networks for program decompilation}.
Nowadays, Artificial intelligence technologies such as deep learning are widely used in NMT, image recognition, and other fields, relying on its powerful learning and expression capabilities to build deep learning-based decompilation methods is a popular research direction~\cite{katz2018using, katz2019towards, fu2019coda, liang2021neutron}. Katz et al.~\cite{katz2018using} first proposed RNN–based method for decompiling binary code snippets, which demonstrated the feasibility of using neural machine translation for decompilation. In order to improve the semantic and grammatical correctness of decompiled code, Katz et al.~\cite{katz2019towards} further proposed a decompilation architecture based on LSTM called TraFix, considering the difference between natural language and programming language. Several approaches, such as preprocess assembly language (TraFix input) and the post-order traversal of C language (TraFix output), have been taken to shorten this difference. Fu et al.~\cite{fu2019coda} proposed an end-to-end code decompilation framework based on several neural networks, different models used for different statement types. Unlike these three exiting works, Our neural translation model \fname is solely based on the self-attention mechanism that can effectively generate high-level PL.

%% file: Conclusion.tex
\vspace{-2mm}
\section{Conclusion}
\label{sec:Conclusion}

In this paper, we propose and implement a novel deep learning-based neural framework for decompilation, named \fnam, which accurately restores the function and semantic information of the target low-level PL code. We also design a novel intermediate language representation named \iname to compensate for information asymmetry between high-level PL and low-level PL. The results on four real-world projects show that the average program accuracy of \fname reaches \cnum{94.41\%}, effectively improving the readability and comprehensibility of the decompiled high-level PL.